\documentclass{aa}
\usepackage[dvips]{graphicx}
\usepackage{times}
\begin{document}
   \title{Wind accretion in binary stars}

   \subtitle{I. Mass accretion ratio}

   \author{T. Nagae\inst{1}, K. Oka\inst{1}, T. Matsuda\inst{1},
          H. Fujiwara\inst{2},
          I. Hachisu\inst{3},
          \and
          H.M.J. Boffin\inst{4,5}}

   \institute{Department of Earth and Planetary Sciences, Kobe University, 
              Rokko-dai 1-1, Nada-ku, Kobe 657-8501, Japan
              \email{nagae@kobe-u.ac.jp, tmatsuda@kobe-u.ac.jp}
         \and
             IBM Japan Ltd., Yamato-shi, Kanagawa 242-8502, Japan
         \and
             Department of Earth Science and Astronomy, College of Arts 
             and Sciences, University of Tokyo, Komaba 3-8-1, Meguro-ku, 
             Tokyo 153-8902, Japan 
             \email{hachisu@chianti.c.u-tokyo.ac.jp}
         \and
             Royal Observatory of Belgium, Av. Circulaire 3,
             1180 Brussels, Belgium
	\and
	     European Southern Observatory,
	     Karl-Schwarzschild-Str. 2, D-85738 Garching, Germany
             \email{hboffin@eso.org}
             }

   \date{Received ; accepted}

   \abstract{
      Three-dimensional hydrodynamic calculations are performed in order to 
      investigate mass transfer in a close binary system, in which one 
      component undergoes mass loss through a wind.
      The mass ratio is assumed to be unity. The radius of the 
      mass-losing star is taken to be about a quarter of the separation between 
      the two stars. Calculations are performed for gases with a ratio of 
      specific heats $\gamma=1.01$ and $5/3$.  Mass loss is assumed to be 
      thermally driven so that the other parameter is the sound speed of the 
      gas on the mass-losing star.

      Here, we focus our attention on two features: flow patterns and mass 
      accretion ratio, which we define as the ratio of the mass accretion 
      rate onto the companion, $\dot{M}_{acc}$, to the mass loss rate 
      from the mass-losing primary star, $\dot{M}_{loss}$.  We characterize 
      the flow by the mean normal velocity of wind on the critical Roche surface 
      of the mass-losing star, $V_R$.  When $V_R<0.4 A\Omega$, 
      where $A$ and $\Omega$ are the separation between the two stars and the 
      angular orbital frequency of the binary, respectively, we obtain 
      Roche-lobe over-flow (RLOF), while for $V_R>0.7 A\Omega$ we observe 
      wind accretion.
      We find very complex flow patterns in between these 
      two extreme cases. We derive an empirical formula of the mass accretion 
      ratio as $0.18 \times 10^{-0.75V_R/A\Omega}$ ~in the low velocity regime 
      and $0.05\,(V_R/A\Omega)^{-4}$ in the high velocity regime.

   }

   \authorrunning{Nagae T. et al.}

   \maketitle

\section{Introduction}

   Wind accretion plays an essential role especially in the evolution of 
   detached binary systems such as symbiotic stars, precursor of
   peculiar red giants, $\zeta$ Auriga stars and massive X-ray binaries.
   In the majority of symbiotic stars, the system contains a hot component -
   believed to be a mass-gaining white dwarf - and a cool component, the 
   mass-losing red giant (e.g., Miko{\l}ajewska \cite{Miko} for a review).  
   Recent studies indicate that many of the red giant 
   components in symbiotic stars do not fill their Roche lobe, i.e., 
   $R_{\rm RG} \la \ell_1 /2$, where $R_{\rm RG}$ is the radius of
   the red giant component and $\ell_1$ is the distance of the inner 
   Lagrangian point from the centre of the red giant 
   (e.g., M\"urset \& Schmid \cite{1999A&AS..137..473M}).  
   This possibly discloses that symbiotic stars are - with probably 
   only the exception of T CrB (e.g., Anupama \& Miko{\l}ajewska 
   \cite{1999A&A...344..177A}) - well detached binary systems. 
   Thus, mass transfer in symbiotic stars seems to be driven by wind, and
   wind mass loss is therefore a key ingredient for triggering symbiotic 
   activity on a white dwarf companion. 
   On the other hand, the mass accretion rates expected in the Bondi-Hoyle 
   picture (Bondi \& Hoyle \cite{1944MNRAS.104..273B}) for such detached 
   binary systems are usually much lower than those required to maintain 
   the symbiotic activity of the systems. 
   For example, Nussbaumer (\cite{Nuss}) pointed out that in most symbiotic 
   stars the typical rates of wind mass loss red giant are in the range  
   $1 \times 10^{-6} M_\odot$ yr$^{-1}$ 
   to $1 \times 10^{-8} M_\odot$ yr$^{-1}$ and that
   the expected accretion efficiency of wind capture is probably  
   no better than one percent.  Hence, the accretion rates 
   associated with most symbiotic white dwarfs are probably 
   in the range of $1 \times 10^{-8} M_\odot$ yr$^{-1}$ to 
   $1 \times 10^{-10} M_\odot$ yr$^{-1}$, which are too low to
   power the typical luminosities of symbiotics ($\ga 1000 L_\odot$)
   unless they are all in a late phase of hydrogen shell-flash (novae)
   on a white dwarf (e.g., Sion \& Ready \cite{1992PASP..104...87S}; 
   Sion \& Starrfield \cite{1994ApJ...421..261S}).
   It will nevertheless be shown in the 
   present paper that the mass accretion rate is smaller than 
   that expected by a simple Bondi-Hoyle picture.

   It has recently been suggested that some of the mass-accreting 
   white dwarfs in symbiotic stars are the progenitors of Type Ia supernovae
   (e.g., Hachisu \& Kato \cite{2001ApJ...558..323H}).  On the other hand, 
   it has long been discussed that the efficiency, the ratio of the mass 
   captured by the white dwarf to the mass lost by the red giant, is very 
   small (say, a few per cent) and not enough to increase the white dwarf mass 
   to the Chandrasekhar mass limit (e.g., Kenyon et al. 
   \cite{1993ApJ...407L..81K}).

   Wind velocities in cool red giants are typically a few times 10 km s$^{-1}$
   and comparable to the orbital velocity in most symbiotic binary systems.
   Therefore, flow patterns are probably something between Roche lobe 
   overflow (RLOF) and  Bondi-Hoyle wind accretion kinds of flows.  
   The efficiency of capture in symbiotics should therefore not be estimated 
   by the Bondi-Hoyle picture but by direct simulations of intermediate, 
   complicated flow patterns.  

   Symbiotic stars are not the only systems in which wind accretion play a 
   significant role. The most adopted model for the formation of peculiar 
   red giants, Barium, CH and S stars, require a binary system containing 
   an Asymptotic Giant Branch star (AGB) transferring mass via its stellar 
   wind to a main sequence companion which becomes polluted in Carbon and 
   s-process elements (see e.g. Boffin \& Jorissen \cite{BJ88}). 
   The AGB then evolve into a white dwarf while the companion will appear 
   as Carbon or s-process rich and, when on the giant branch, as a peculiar 
   red giant. Boffin \& Zacs (\cite{BZ94}) have shown that an accretion 
   efficiency of only a few percent is enough to explain present Barium stars.
   Objects related to these are the extrinsic S stars which also show 
   symbiotic activity (e.g. Carquillat et al. \cite{Carq98}) and bipolar 
   planetary nebulae (e.g. Mastrodemos \& Morris \cite{MM98}).

   $\zeta$ Aurigae systems are another kind of system where wind accretion 
   plays a role. In these eclipsing double-lined binaries, a main sequence 
   star has an accretion wake produced by the wind of its K supergiant 
   companion. The eponymous system, $\zeta$ Aur, is a binary with a 972 days 
   orbital period and containing a 5.8 M$_\odot$ K4 Ib and a 4.8  M$_\odot$ K 
   B5 V star (Bennett et al. \cite{Benn96}).

   Finally, accretion wakes are also reported for high mass X-ray binaries 
   (HMXB). These systems consist of a compact object, neutron star or black 
   hole, orbiting a massive OB primary star which has a strong stellar wind. 
   The X-ray emission is believed to be due to accretion of matter on the 
   compact companion.

   There have been already many numerical studies of stellar wind in a close 
   binary system. Biermann (\cite{1971A&A....10..205B}) computed 
   two-dimensional stellar wind using a characteristic method. Sorensen, 
   Matsuda \& Sakurai (\cite{1975Ap&SS..33..465S}) performed two-dimensional 
   finite difference calculation of stellar wind emitted from a Roche lobe 
   filling secondary. They used the Fluid in Cell (FLIC) method with first 
   order of accuracy and a Cartesian grid. They computed Roche lobe overflow 
   (RLOF) as well. Sawada, Hachisu \& Matsuda (\cite{1984MNRAS.206..673S}) 
   calculated two-dimensional stellar wind from a contact binary using 
   a Beam-Warming time implicit finite difference scheme and a generalized 
   curvilinear coordinate. Their main goal was to compute the angular 
   momentum loss rate 
   from the system, which is an important factor to define the evolution of 
   the binary system. Sawada, Matsuda \& Hachisu (\cite{1986MNRAS.221..679S}) 
   computed two-dimensional stellar wind from a semi-detached binary system 
   using the Osher upwind scheme and a generalized curvilinear coordinate. 
   They observed a transition from RLOF to stellar wind by increasing the wind 
   speed on the mass-losing star. The present study is a three-dimensional 
   version of their study.  Matsuda, Inoue \& Sawada, 
   (\cite{1987MNRAS.226..785M}) also conducted a similar study and 
   discovered a flip-flop instability of the bow shock formed around a 
   compact mass accreting object (see also Boffin \& Anzer, \cite{BA94}, 
   and on the stability issue, Foglizzo \& Ruffert, 1997 and 1999, and 
   Pogolerov, Ohsugi \& Matsuda, 2000). 
   However, it was found that the phenomenon was characteristic to 
   two-dimensional case (Matsuda et al., \cite{1992MNRAS.255..183M}, 
   Ruffert, \cite{Ruffert96}). In the present work we perform 
   three-dimensional calculation in which we do not observe the flip-flop 
   instability.

   These simulations concerned the case of a compact object moving in a 
   wind and were mostly aimed at estimating the validity of the Bondi-Hoyle 
   accretion rate. Binary effects were partially simulated by including 
   velocity or density gradient in the wind. There are a few simulations 
   however which simulate in full mass transfer by wind in a binary system 
   (Blondin et al. 1991, Theuns \& Jorissen 1993, Blondin \& Woo 1995, 
   Theuns et al. 1996, Walder 1997, Mastrodemos \& Morris 1998, Dumm et al. 
   2000, Boroson et al. 2001, 
   Gawryszczak, Miko{\l}ajewska, \& R{\' o}{\. z}yczka 2002). 
   Theuns \& Jorissen (\cite{TJ93}) and Theuns, 
   Boffin \& Jorissen (\cite{TBJ96}, TBJ96 in the following) performed 
   three-dimensional hydrodynamic simulations of a 3 M$_\odot$ AGB 
   transferring mass through its stellar wind to a 1.5 M$_\odot$ main 
   sequence companion, in order to test the wind accretion model for the 
   origin of peculiar red giants. They performed simulations using a 
   polytropic equation of state with $\gamma =$1., 1.1 and 1.5 and found mass 
   accretion ratios  (that is, mass accretion rate/mass loss rate)
   of 1-2\% for the $\gamma =1.5$ case and 8 \% for the 
   other models, i.e. about ten times smaller than the theoretical Bondi-Hoyle 
   estimate. They also observed that this value is dependent on resolution. 

   Walder (1997) presented simulations of wind accretion in well separated 
   binaries for three different cases: a HMXB, $\zeta$ Aur and a barium star 
   progenitor. He obtains mass accretion ratio 
   (mass accretion rate/mass loss rate) of 0.6, 3 and 6 \%, 
   respectively. Walder \& Folini (2000) also presented a nice review of 
   wind dynamics in symbiotics binaries, with an emphasis on the influence
   of the radiation field of the accretor.

   More recently, Dumm et al. (\cite{Dumm00}) presented three-dimensional 
   Eulerian isothermal simulations in order to represent the symbiotic system 
   RW Hya which seems to present an accretion wake, as indicated by the 
   reduced UV flux observed at phase $\Phi = 0.78$. They found that 6\% of 
   the M-giant wind is captured by the companion. 

   In this paper we perform three-dimensional numerical simulations of mass 
   transfer by wind in a binary system with a mass ratio of unity. The full 
   gravitational forces of the two components are taken into account. 
   This is therefore different from the e.g. TBJ96 or Dumm et al. (2000) 
   studies, where in order to account for the not very well known acceleration 
   mechanism of the wind, the gravitational force of the primary is partially 
   or totally reduced.

   Here, we will concentrate on the derivation of the mass accretion 
   ratio. 
   We will leave for a following paper the physical discussion of the 
   flow, including a discussion of angular momentum loss from the system.

\section{Assumptions and numerical method}
\subsection{Model}

   Consider a detached binary system of equal mass: one larger star 
   blows stellar wind which the other, more compact, component partly 
   accretes. 
   We normalize the length by the separation of the two stars, $A$, and the 
   time by $1/\Omega$, where $\Omega$ is the angular velocity of the binary 
   system. The surface of the mass-losing star is assumed to be an 
   equi-potential surface, whose mean radius is 0.25. The radius 
   of the mass accreting star is assumed to be 0.015 or smaller. 

   From an OB star, stellar wind is mainly accelerated by line 
   absorption of UV radiation. In symbiotic stars or in precursors of
   peculiar red giants (PPRGs), the stellar wind mechanism  
   from the cool star is poorly known, and may imply radiation onto dust 
   grains. One needs however to still bring the matter far enough away from 
   the star so that dust grains can form. In more evolved red giants, 
   this may be done through the effect of pulsation. It is however still not 
   clear what 
   the mechanism could be in non-pulsating red giants, as they seem to exist
   in some symbiotic or extrinsic S stars.

   It is therefore clear that simplifying assumption have to be made. 
   In the present study, we will thus simply assume that 
   the wind is generated by thermal pressure, like in the solar wind. 
   More complicated wind structures will be studied in a forthcoming paper.

   We assume that the gas is an ideal one and is characterized by the ratio of 
   specific heats, $\gamma$. In the present work we consider two extreme 
   cases : $\gamma=5/3$ and $\gamma=1.01$. In the latter case, the gas is 
   almost isothermal except where there are shocks. This isothermality may 
   be explained as an outcome of good thermal conduction. We neglect other 
   complex effects like magnetic fields, radiation and viscosity 
   (except numerical one).

   We have carried out our simulations for the case of a mass ratio unity.
   Although this may not be representative for PPRGs, symbiotic stars or 
   HMXBs, this should not change very much our results.

\subsection{Method of calculation}

   We solve the three-dimensional Euler equations using a finite volume 
   scheme. As  Riemann solver, we use the SFS scheme and we apply the MUSCL 
   method to interpolate physical variables in a cell; the details 
   on the scheme is described in Makita, Miyawaki \& Matsuda 
   (\cite{2000MNRAS.316..906M}), and RLOF simulations using the same method are 
   given in Fujiwara et al. (\cite{2001PThPh.106..729F}).

\begin{figure}[htbp]
      \centering
      \includegraphics[width=8.8cm]{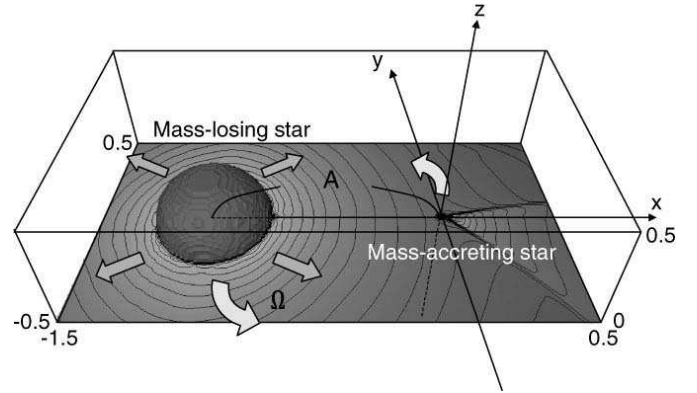}
      \caption{Computational region.}
               \label{CR}
   \end{figure}

   Calculations are done in the rotating frame of a binary and include the
   gravity of the two stars, the centrifugal and Coriolis forces.
   We use a Cartesian coordinate with the origin at the mass accreting star. 
   The computational region is $-1.5<x<0.5$, $-0.5<y<0.5$, $0<z<0.5$ and
   we assume symmetry around the orbital plane (Fig. \ref{CR}).

   The region is divided into $201 \times 101 \times 51$ cells. 
   The mass accreting companion star is represented by a cubic hole with 
   cells of $3\times3 \times 3$, including the lower half of the hole 
   under the orbital plane. Its physical dimension is 
   $0.03 \times 0.03 \times0.03$ in our units.  

   In order to study in details the structure of the flow near the mass 
   accreting object, we introduce a three levels nested grid. At the 
   second level, the region $-0.25<x<0.25$, $-0.25<y<0.25$ and $0<z<0.25$ is 
   further divided into $102 \times 101 \times 51$ cells. At the third 
   level, the region $-0.125<x<0.125$, $-0.125<y<0.125$ and $0<z<0.125$ is 
   once more divided into $104 \times 104 \times 54$ cells. The cell size 
   of the third level is thus 1/4 of the original first level cell. The hole 
   of $3 \times 3 \times 3$, in the first level, is represented 
   by a hole with $12 \times 12 \times 12$ cells in the third level. 

   A typical symbiotic star, $\zeta$ Auriga or PPRG has an orbital period of 
   a few years and a separation of a few AU. In this case, the cell size of 
   0.03 is of the order or even larger than a main sequence star. 
   This is thus applicable for $\zeta$ Aurigae or PPRGs. For S or symbiotic
   stars, we should simulate the case of an accreting white dwarf. This would 
   require a hole about 100 times smaller. This is well beyond present days 
   numerical capabilities. This is also true if one would like to simulate 
   the compact accreting object in HMXBs. 

   One should note however that the size of our accreting object is still 
   smaller than the one adopted by e.g. TBJ96. They have shown that the mass 
   accretion rate decreases with decreasing accreting size and our value is 
   similar to their highest resolution model. We thus also consider a case 
   of a hole having a size of $3\times 3 \times 3$ cells
   (i.e. of physical dimensions $\simeq 0.01 \times 0.01 \times 0.01$ at the 
   third level in order to investigate the effect of the size of the mass 
   accreting object on the results.

\subsection{Boundary conditions and initial condition}

   The outer parts of the mass-losing star is filled by a gas of 
   density $\rho=1$ and sound speed $c_s$. If the pressure inside the 
   mass-losing star is larger than that outside the mass-losing star, 
   gas is ejected from the surface. The velocity of the ejected gas is 
   computed by solving a Riemann problem between the two states. 

   At $t=0$ the computational region except the mass-losing star is filled 
   by a tenuous gas with density $\rho_0=10^{-5}$, velocity $u_0=v_0=w_0=0$ 
   and pressure $p_0=10^{-4}/\gamma$. The computational region is gradually 
   filled with the gas ejected from the mass-losing star. The outside 
   of the computational region is filled by the initial gas all the time. 
   Mass outflow/inflow from the outer boundary can occur; the amount of 
   outflow/inflow is also calculated by solving Riemann problems between two 
   states. This (outside) boundary condition, which we call the ambient 
   condition, insures the stability of the computation. 
   The inside of the mass accreting hole is almost vacuum and the gas 
   approaching the hole is absorbed.

   We start the calculation at $t=0$ and terminate it when the system becomes 
   steady. In the wind cases half an orbital period is enough to reach 
   a steady state, but in the RLOF cases we have to calculate at least 
   a few orbital periods.

\section{Numerical results}
\subsection{Parameters}

   As was described earlier, we adopt $\gamma$ and the dimensionless sound 
   speed of gas in the mass-losing star, $c_s$, as the model parameters. 
   In the present work, we consider the cases of $\gamma=1.01$ and 
   $\gamma=5/3$. When $c_s$ is less than some critical value, we expect 
   RLOF type flows, while in the other cases, we expect stellar wind kind 
   of flows. As is shown in Table \ref{Tab:Simu}, we study the cases of 
   $c_s=0.5-2.5$ for $\gamma=1.01$, and $c_s=0.75-2.5$ for $\gamma=5/3$. 

   In PPRGs and symbiotic stars, the sound speed at the surface of the 
   mass-losing giant is of the order of 5-20 kms$^{-1}$ while the orbital 
   velocity is of the order of 10-40 kms$^{-1}$. In $\zeta$ Aurigae, both 
   these values are around 40-100 kms$^{-1}$, while in HMXBs, the wind speed
   is of the order of a few thousands of kms$^{-1}$ while the orbital 
   velocity is several hundreds of kms$^{-1}$.

   The parameter $c_s$ has no physical significance, because the mean radius, 
   0.25, of the mass losing star is chosen arbitrarily for numerical 
   purposes. Using the results of our numerical simulations, we therefore 
   calculate the mean value of the vertical component 
   of the gas speed on the critical Roche surface, $V_R$. We classify the flow 
   pattern using $V_R$, which is shown in Table~\ref{Tab:Simu} as well.
   It has to be emphasized that  $V_R$ is not constant along 
   the critical Roche surface, especially in the Roche lobe type of flows. For
   these, it may be better to use $c_s$ as a parameter. For the wind accretion
   type of flow, which are more isotropic,  $V_R$ is more appropriate.

\begin{table*}
  \centering
  \caption[]{Model parameters. The parameters $c_s$, $V_R$ and $f$ 
           are the sound speed on the surface of the mass losing star,
           the normal velocity at the critical Roche surface and the 
           mass accretion ratio.}
  \label{Tab:Simu}

  \begin{tabular}{ccccccccccccc}
  \hline
  \hline
  \noalign{\smallskip}
        & & \multicolumn{3}{c}{$\gamma=5/3$} & &
            \multicolumn{3}{c}{$\gamma=1.01$} & &
            \multicolumn{3}{c}{Nested Grid ($\gamma=1.01$)} \\
  $c_s$ & &  Case & $V_R$ & $f$ & &
             Case & $V_R$ & $f$ & &
             Case & $V_R$ & $f$ \\ 
  \hline 
  \noalign{\smallskip}
   0.5  & & \multicolumn{3}{c}{-----------} & & 
            a & 0.06 & $1.7\times10^{-1}$ & &
            \multicolumn{3}{c}{-----------} \\
   0.75 & & B & 0.01 & $3.4\times10^{-1}$ & &
            b & 0.46 & $9.2\times10^{-2}$ & &
            b$^{\prime}$ & 0.46 & $1.1\times10^{-1}$ \\
   0.9  & & C & 0.03 & $1.0\times10^{-1}$ & & 
            c & 1.09 & $2.4\times10^{-2}$ & &
            c$^{\prime}$ & 1.09 &$2.2\times10^{-2}$ \\
   1.0  & & D & 0.12 & $1.9\times10^{-2}$ & & 
            d & 1.41 & $1.1\times10^{-2}$ & &
            d$^{\prime}$ & 1.41 & $1.2\times10^{-2}$ \\
   1.2  & & E & 0.68 & $3.3\times10^{-2}$ & & 
            e & 1.95 & $3.8\times10^{-3}$ & &
            e$^{\prime}$ & 1.94 & $3.9\times10^{-3}$ \\
   1.4  & & F & 1.49 & $1.1\times10^{-2}$ & & 
            f & 2.42 & $1.6\times10^{-3}$ & &
            f$^{\prime}$ & 2.41 & $1.8\times10^{-3}$ \\
   1.6  & & G & 1.86 & $7.4\times10^{-3}$ & & 
            g & 2.86 & $6.4\times10^{-4}$ & &
            g$^{\prime}$ & 2.85 & $5.4\times10^{-4}$ \\
   1.8  & & H & 2.20 & $3.6\times10^{-3}$ & & 
            h & 3.29 & $4.7\times10^{-4}$ & & 
            \multicolumn{3}{c}{-----------} \\
   2.0  & & I & 2.52 & $2.1\times10^{-3}$ & & 
            i & 3.71 & $3.6\times10^{-4}$ & &
            i$^{\prime}$ & 3.69 & $3.3\times10^{-4}$ \\
   2.5  & & J & 3.29 & $7.9\times10^{-4}$ & & 
            j & 4.75 & $2.4\times10^{-4}$ & &
            j$^{\prime}$ & 4.71 & $2.3\times10^{-4}$ \\ 
\hline
  \end{tabular}
\end{table*}

\subsection{Roche lobe overflow}

\begin{figure}[htbp]
   \centering
   \includegraphics{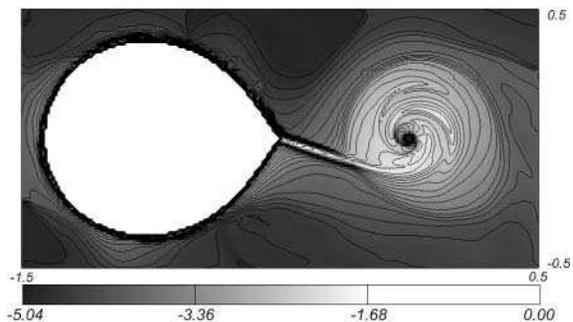}
   \caption{Density contours in the orbital plane of the typical RLOF
            case, but in a semi-detached binary system 
            ($\gamma=1.01$, $c_s=0.02$).}
            \label{RLOF1}
   \end{figure}

\begin{figure}[htbp]
   \centering
   \includegraphics{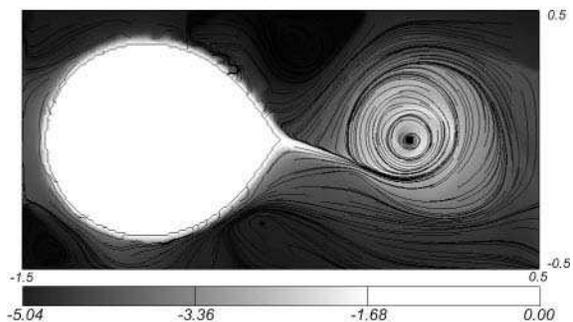}
   \caption{Streamlines and Mach number counters of the typical RLOF case
            ($\gamma=1.01$, $c_s=0.02$).}
            \label{RLOF2}
   \end{figure}

   In a typical Roche lobe overflow (RLOF), gas on the surface of the 
   companion flows through the L1 point and forms a narrow L1 stream. The L1 
   stream deflects its motion to the negative $y$ direction in the present 
   configuration due to the Colioris force, and then circulates 
   counterclockwise about the mass-accreting star to form an accretion 
   disc. 
   We show an example of such typical ROLF flow in the case of $\gamma =1.01$ 
in Fig.~\ref{RLOF1} and \ref{RLOF2}. Here we assumed the star to fill its
Roche lobe, so this is not the main aim of this paper. They will serve as
comparison for the B-D and $a$ cases discussed below.

   If the temperature of the surface of the companion star, $c_s$, 
   is raised, the L1 stream becomes thicker (Fujiwara et al., 
   \cite{2001PThPh.106..729F}). We define the flow to be of the RLOF type if 
   the flow circulates around the accreting object in counterclockwise 
   direction. 

   If we raise $c_s$ further, the gas escapes from the mass losing star in all 
   directions, and part of the gas rotates in a clockwise direction around the 
   companion. These two flows, rotating counterclockwise and clockwise, 
   collide behind (i.e. to the right) the accreting object to form 
   a bow shock. This is a typical wind accretion flow. 

   In the B-D cases  for $\gamma=5/3$ and the $a$ case for $\gamma=1.01$, 
   we see from Table~\ref{Tab:Simu} that $V_R < 0.4$, i.e. $V_R$ is smaller 
   than the escape velocity. The flow is of the RLOF type.
   Similar types of flows have been obtained by other authors, 
   e.g. Bisikalo et al. (1998).

\begin{figure}[htbp]
   \centering
   \includegraphics{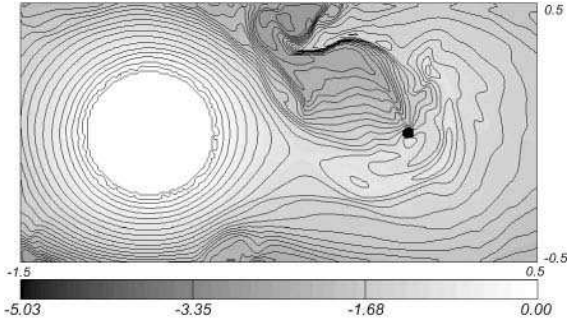}
   \caption{Density contours in the orbital plane of the RLOF type flow of
            Model B ($\gamma=5/3$, $V_R=0.01$). The left circle shows the 
            mass losing companion star with a radius of 0.25, and the 
            right black dot indicates the mass accreting object with 
            a radius of 0.015.}
            \label{RLOF3}
   \end{figure}

\begin{figure}[htbp]
   \centering
   \includegraphics{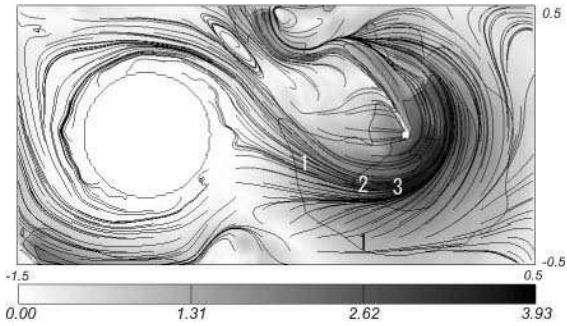}
   \caption{Streamlines and  Mach number contours in the orbital 
            plane of the same model B as in Fig.~\ref{RLOF1}. The Mach 
            number contours are represented by the shade, and the numbers 
            in the figure show typical Mach numbers. Streamlines are 
            generated from points separated by the same distance. We observe 
            that the flow rotates counterclockwise around the companion, 
            which is a typical feature of RLOF type flows.}
            \label{RLOF4}
    \end{figure}

   Figure~\ref{RLOF3} shows the density contours for model B 
   ($\gamma=5/3$, $V_R=0.01$), and Figure \ref{RLOF4} depicts the streamlines 
   for the same case. 
   In this case, since $\gamma=5/3$, the temperature 
   of the disc is so hot that the mass accretion ratio $f$, which is defined 
   in Sect.~\ref{sec:MAR}, is not 1 but 0.34. Spiral shocks, typical of 
   RLOF flows (Makita et al. \cite{2000MNRAS.316..906M}, Matsuda et al. 
   \cite{2000Ap&SS.274..259M}, Fujiwara et al. \cite{2001PThPh.106..729F}) 
   are not seen in this case.

   One should note that contrarily to the impression one might have 
when just looking at the figures, we have checked that there is no inflow 
from the outer boundaries in these simulations.

\begin{figure}[htbp]
   \centering
   \includegraphics{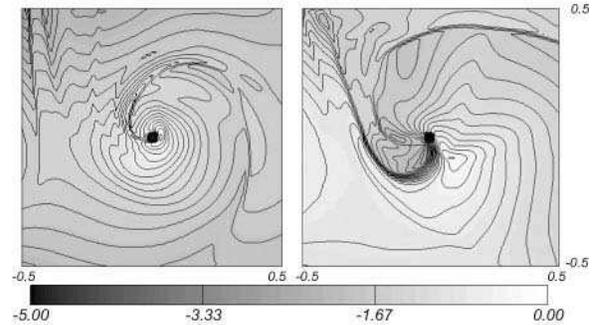}
   \caption{Comparison of two RLOF type flows. Density contours in the 
            orbital plane are shown. Left: Model a ($\gamma=1.01$, 
            $V_R=0.06$); Right: Model D ($\gamma=5/3$, $V_R=0.12$). 
            We observe spiral shocks which are typical to RLOF.}
            \label{Comparison_gamma1}
    \end{figure}

   Figure \ref{Comparison_gamma1} shows the comparison of model $a$ 
   ($\gamma=1.01$, $V_R=0.06$) and D ($\gamma=5/3$, $V_R=0.12$). 
   We observe spiral shocks, although with a very much distorted shape.

\subsection{Intermediate flow}

\begin{figure}[htbp]
   \centering
   \includegraphics{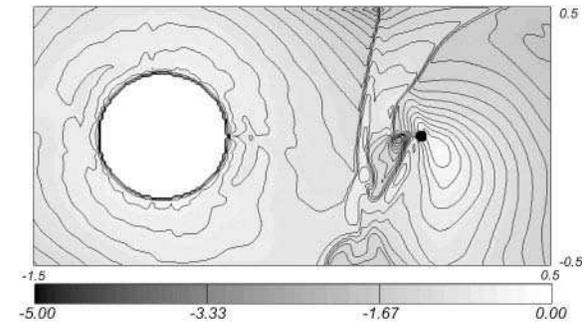}
   \caption{Density contours in the orbital plane of the model E 
            ($\gamma=5/3$, $V_R=0.68$). We observe very complicated shock 
            patterns.}
            \label{IMF1}
   \end{figure}

\begin{figure}[htbp]
   \centering
   \includegraphics{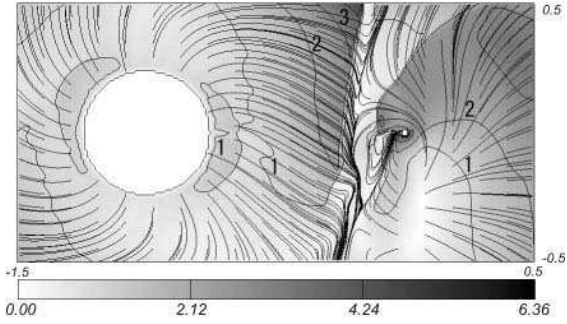}
   \caption{Streamlines and Mach number contours in the orbital 
            plane of the same model E as in Fig. \ref{IMF1}.}
            \label{IMF2}
   \end{figure}

   In the cases E, b and b$^{\prime}$, we observe that $0.4 < V_R < 0.7$. 
   In these cases, the flow is not a typical RLOF or a wind accretion flow, 
   but rather a very complicated intermediate one. Figure \ref{IMF1} shows 
   the density contours of model E, and Fig.~\ref{IMF2} depicts the 
   streamlines of the same model. 

   \begin{figure}[htbp]
   \centering
   \includegraphics{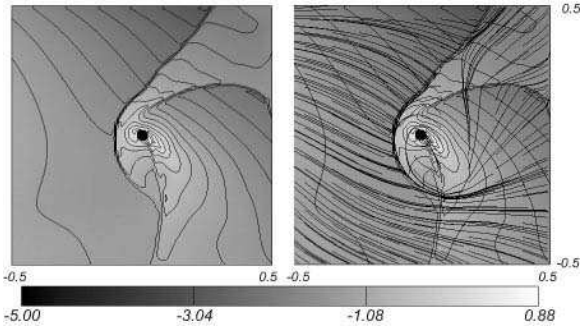}
   \caption{Density contours and stream lines in the orbital plane of 
            the model b ($\gamma=1.01$, $V_R=0.46$).}
            \label{Model_b}
   \end{figure}

   Figure \ref{Model_b} shows the density contours and the streamlines of the case b. 
   We find that the flow pattern and the shock structure are very complicated in these intermediate flow. 

   We will now speculate on the possible cause of the complicated structure of 
   the intermediate flow. Gas 
   flows from the mass-losing star (left sphere) towards the mass accreting 
   object (right small sphere). The lower part of the flow experiences a 
   shock, and is decelerated to subsonic speed. A small accretion disc is 
   formed around the companion, but some part of the gas is accelerated to 
   supersonic velocity again. It collides with the gas coming from the 
   positive $y$ side and forms a bow shock. The orientation of the bow shock 
   is rotated almost 90 degrees counterclockwise.

   \begin{figure}[htbp]
   \centering
   \includegraphics[width=8.0cm]{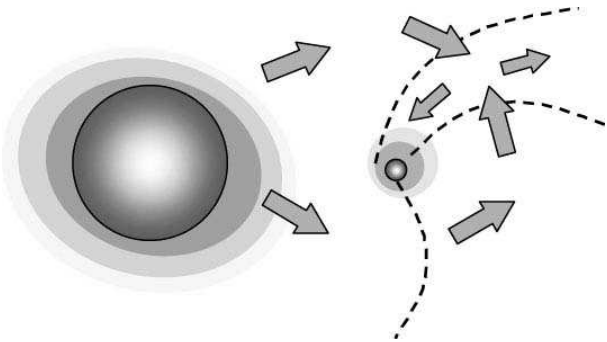}
   \caption{A schematic diagram of an intermediate flow model E.}
            \label{IMF_MODEL}
   \end{figure}
\subsection{Wind accretion flow}

\begin{figure}[htbp]
   \centering
   \includegraphics[width=8.0cm]{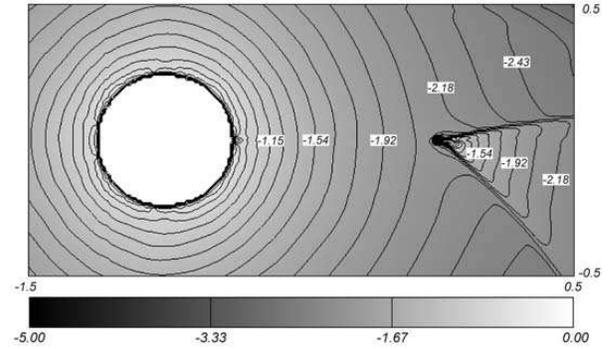}
   \caption{Density contours in the orbital plane of a typical wind 
            accretion flow model g ($\gamma=1.01$, $V_R=2.86$). We observe 
            a conical bow shock attached to the accreting object, which is 
            a typical feature of the wind accretion flow.}
            \label{WAF1}
   \end{figure}

\begin{figure}[htbp]
   \centering
   \includegraphics{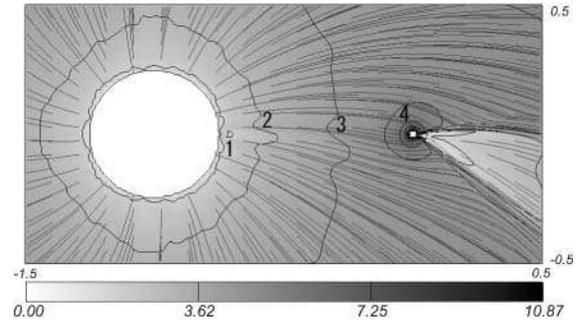}
   \caption{Stream lines of the model g as in Fig. \ref{WAF1}.
            Numbers are magnitude of normalized velocity.}
            \label{WAF2}
   \end{figure}

   For the cases with $V_R > 0.7$, which are all the other models than stated 
   above, the gas ejected from the mass-losing star is accelerated by 
   thermal pressure. Almost all gas escapes from the system and only 
   a small part of gas is accreted by the companion. We find a typical 
   conical bow shock attached to the accreting object. Figures 
   \ref{WAF1} and \ref{WAF2} show the density distribution, 
   the streamlines and the magnitude of the velocity for model g 
   ($\gamma=1.01$, $V_R=2.86$). 

\subsection{Calculations based on the nested grid}

\begin{figure}[htbp]
   \centering
   \includegraphics{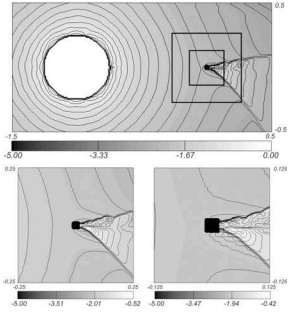}
   \caption{Density contours of the model g$^{\prime}$ 
            ($\gamma=1.01$, $V_R=2.85$) calculated on the nested grid. 
            Two blow-ups close to the accreting star are shown. 
            Compare the present figure with Fig.~\ref{WAF1}. We observe 
            that the bow shock is slightly jaggy compared with Fig.10. 
            This is due to the reduced numerical viscosity.}
            \label{Nest}
   \end{figure}

   In order to test the effect of resolution near the mass accreting 
   object, we calculate the flow using the nested grid described above. Figure 
   \ref{Nest} shows the density contours for model g$^{\prime}$. If we compare 
   the result of model g$^{\prime}$ with that of model g, we can see a slight 
   oscillation on the bow shock for model g$^{\prime}$. Nevertheless, 
   the essential features are unchanged except for the mass accretion 
   efficiency.

\subsection{Mass accretion ratio}
\label{sec:MAR}

\begin{figure}[htbp]
   \centering
   \includegraphics[width=8.8cm]{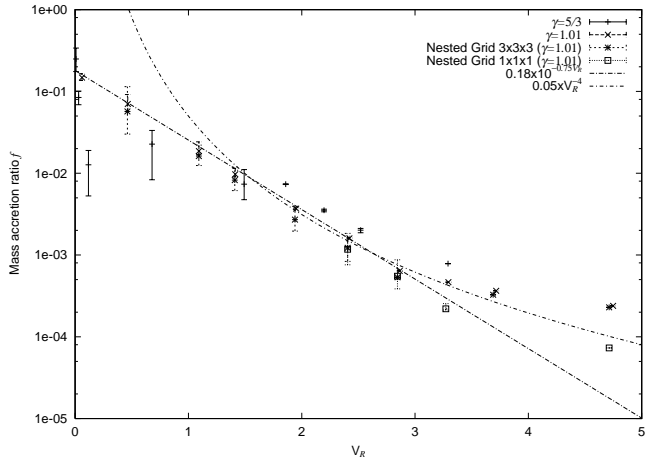}
   \caption{Mass accretion ratio, defined by the ratio of the mass 
            accretion rate to the mass loss rate, as a function of 
            $V_R$. The horizontal axis is $V_R$ and the vertical one is 
            the mass accretion ratio. The dash-dotted line 
            is our empirical formula for the mass accretion ratio: 
            $0.18 \times 10^{-0.75V_R}$, 
	    while the upper line is the fit valid for large $V_R$. }
            \label{Accretion_Rate}
   \end{figure}

   We calculate the mass accretion ratio $f$, which is defined as the ratio 
   of the mass accreting rate onto the companion, $\dot{M}_{acc}$, 
   to the mass loss rate from the mass losing star, $\dot{M}_{loss}$. 
   The mass accretion ratio as a function of $V_R$ is shown in Fig. 
   \ref{Accretion_Rate}. The dash-dotted line is an empirical formula 
   $f=0.18 \times 10^{-0.75~V_R}$, which is valid in the range $V_R< 2.5$.

   In typical wind case we found that $f$ is less than 1\%, and decreases 
   further with increasing $V_R$ although it seems to saturate (but see below). 
   In the intermediate flow regime, $f$ is about 
   a few \% to 10\%. For RLOF $f$ is greater than 10 \% for $\gamma=1.01$.
   If we use the above empirical formula to the limit of $V_R=0$, we have 
   $f=0.18$. Of course, this is not necessary true and $f=1$ is also possible. 
   Although a direct comparison is not possible, our values are in good 
   agreement with the results obtained by TBJ96, Walder (\cite{Walder97}), 
   Mastrodemos \& Morris (1998) or Dumm et al. (2000).

   The curve of the calculated accretion ratio based on the coarse grid has 
   a kink about $V_R=2.6$. This may be due to the fact that the size of 
   the accreting cubic box may be comparable to or larger than the accretion 
   radius. In order to see the effect, we reduced the size of the inner box. 
   We found that the position of the kink is shifted towards larger $V_R$, 
   and the accretion ratio is reduced. This again is in agreement with the
   analysis done by TBJ96. 

   For high wind velocity, we expect the Hoyle-Lyttleton theory of wind accretion 
   (Hoyle \& Lyttleton \cite{HL}), that is the case where pressure force is neglected, to be applicable. 
   If it is the case, then the mass accretion rate $\dot M_{acc}$ must be 
   proportional to the square of the mass accretion radius, $2GM/V_{\inf}^2$ - with $V_{\inf}$ being the wind velocity at infinity, 
   and so we expect that $f \sim V_{\inf}^{-4}$. Assuming a spherical symmetric 
   wind around the mass-losing component, we have
   \begin{equation}
     \dot M_{\rm wind} = 4 \pi A^2 \rho_A V_A,
     \label{wind_loss}
   \end{equation}
   near the mass-accreting component.  For a high wind velocity, we may
   expect that the accretion is determined by Hoyle \& Lyttleton's
   formula, i.e.,
   \begin{equation}
     \dot M_{\rm acc} = \pi r_a^2 \rho_A V_A,
     \label{hoyle_accretion}
   \end{equation}
   and the accretion radius $r_a$ is given by
   \begin{equation}
     r_a = {{2 G M_1} \over {V_A^2}},
     \label{accretion_radius}
   \end{equation}
   where $\rho_A$ is the mass density, $V_A$ is the wind velocity 
    measured on the rotating frame, 
   both
   at the mass accreting component, $M_1$ is the mass of the mass-accreting
   component.  Therefore, the accretion ratio is calculated from
   \begin{equation}
     f = {{\dot M_{\rm acc}} \over {\dot M_{\rm wind}}}
     \approx \left( {{V_A} \over {A \Omega}} \right)^{-4}
     \left({{M_1} \over {M_1+M_2}} \right)^2.
     \label{accretion_ratio}
   \end{equation}
   Here, we use Kepler's relation of $A^3 \Omega^2 = G(M_1+M_2)$ and
   $M_2$ is the mass of the mass-losing component.
   For the present case of equal mass, the accretion ratio becomes
   \begin{equation}
     f = 0.25 \times \left( {{V_A} \over {A \Omega}} \right)^{-4}.
     \label{equal_mass_ratio}
   \end{equation}
               
   Note that we checked that the wind velocity does not change by a noticeable amount between the
   Roche lobe surface of the companion to the position of the primary. Hence, we can put $V_A = V_R$ in the above equation.

\begin{figure}[htbp]
   \centering
   \includegraphics[width=8.8cm]{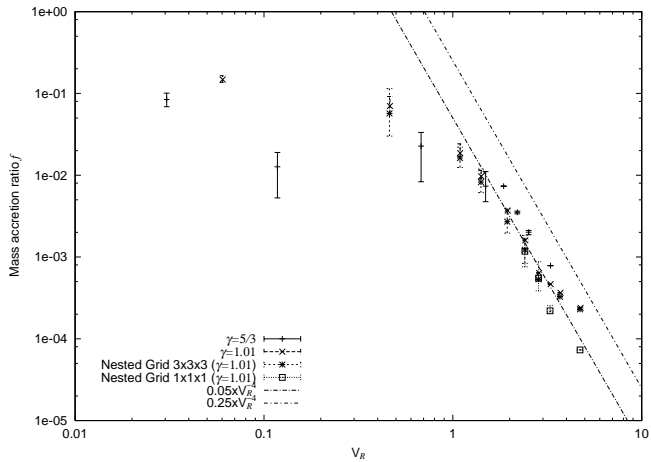}
   \caption{Log-log plot of the Mass accretion ratio as a function of the 
            wind velocity at the position of the Roche lobe. The 
            upper line shows a simple Hoyle-Lyttleton formula as given 
	    by Eq.~(\ref{equal_mass_ratio}). 
	    The lower line shows another 
            empirical formula $f = 5.0\times10^{-3} V_R^{-4}$.}
\label{log-log}
   \end{figure}

   In Fig.~\ref{log-log} we show the accretion ratio as a function of 
   $\log V_R$. The upper line shows the simple Hoyle-Lyttleton 
   formula as given by Eq.~(\ref{equal_mass_ratio}). The lower line 
   shows another experimental formula  $0.05~V_R^{-4}$, which 
   agrees well with the calculated results for higher $V_R$. It appears thus 
   that the calculated mass accretion efficiency is  about 20 \% 
   of the predicted one. 
   Such a comparison is however not fair for at least three reasons. 
   First, because we take into account pressure, one should not use 
   Hoyle-Lyttleton relation but rather the Bondi-Hoyle formula, and include 
   the sound speed. 
   Second, because our wind is thermally driven and we are in a binary 
   system, it is not obvious at all to determine which wind velocity should be 
   inserted in the equation. 
   In the Hoyle-Lyttleton formalism, the velocity which is inserted is the 
   upstream wind velocity at infinity. This is clearly meaningless in a close 
   binary where the wind may not have reached its terminal velocity while 
   close to the companion. 
   Third, our wind is a diverging flow from the mass losing star, which contradicts 
the parallel upstream flow assumed in the Hoyle-Lyttleton or Bondi-Hoyle treatment.
   Therefore the use of the Hoyle-Lyttleton or Bondi-Hoyle formalism can 
   be justified only qualitatively.

\subsection{Time variation}

   We monitored the time variation of the mass accretion ratio. For 
   intermediate flows the amplitude of time variation is about 30\% for 
   $\gamma=1.01$ and it is about 70\% for $\gamma=5/3$. This trend that the 
   amplitude of oscillation is larger for larger $\gamma$ is applicable to 
   other cases. For larger $V_R$ cases the oscillation is negligible. In Fig. 
   \ref{Accretion_Rate}, the error bars show the amplitude of the oscillation 
   of the mass accretion ratio. As was discussed above, the amplitude is 
   larger for the finer grid because of the reduction of the numerical 
   viscosity. By decreasing the size of the accreting box, we increase the 
   range in $V_R$ showing time variability. We Fourier analyzed the time 
   sequence of mass accretion and did not find any typical frequency.

\section{Conclusion}

\begin{enumerate}
   \item The flow pattern is classified by means of the normal speed of gas 
   on the critical Roche surface, $V_R$. For $V_R<0.4A\Omega$, we have a RLOF 
   type flow, while for $V_R>0.7A\Omega$, wind accretion flows are realized. 
   In the intermediate parameter range, $0.4A\Omega<V_R<0.7A\Omega$, we 
   observe very complex flows.

   \item We construct an 
    empirical formula for the mass accretion ratio given by 
    $f = 0.18\times 10^{-0.75 V_R/A\Omega}$, valid for low $V_R$.

   \item The calculated mass accretion ratio shows a kink at 
   $V_R \sim 2.6A\Omega$. By using the nested grid, we found that this is 
   because the assumed size of the mass-accreting box is comparable to or 
   larger than the accretion radius. By reducing the size of the 
   mass-accreting box for the nested grid, the position of the kink is 
   shifted to larger $V_R$. 

  \item For larger $V_R$ we obtain another empirical formula for the mass
   accretion ratio  $f = 0.05(V_R/A\Omega)^{-4}$.

   \item For typical wind accretion flow, the results based on the coarse 
   grid and those of the nested grid are essentially the same except the mass 
   accretion ratio described above. For intermediate flow, the computed flow 
   pattern depends on the resolution slightly. The amplitude of the 
   oscillation of the flow increases with increased resolution because of 
   reduced numerical viscosity.
\end{enumerate}

\begin{acknowledgements}

   We thank the anonymous referees for useful comments and remarks.
   Calculations were performed on the Origin 3800 at the information 
   processing centre of Kobe University and on the NEC SX-5 at the Yukawa 
   Institute of Kyoto University. 
   K.O. was supported by the Research Fellowships of the Japan Society for
   Promotion of Science for Young Scientists.
   T.M. was supported by the grant in aid 
   for scientific research of the Japan Society of Promotion of Science 
   (13640241) and I.H. by (11640226). This work was supported by "The 21st
   Century COE Program of Origin and Evolution of Planetary Systems" in 
   Ministry of Education, Culture, Sports, Science and Technology (MEXT).

\end{acknowledgements}


\begin{thebibliography}{}

  \bibitem[1999]{1999A&A...344..177A} Anupama, G. C., \& Miko{\l}ajewska, J. 
     1999, \aap, 344, 177

  \bibitem[1996]{Benn96} Bennett, P. D., Harper, G. M., Brown, A., \&
     Hummel, C. A. 1996, \apj, 471, 454

  \bibitem[1971]{1971A&A....10..205B} Biermann, P. 1971, \aap, 10, 205

  \bibitem[1991]{Bl91} Blondin, J. M., Stevens, I. R., \& Kallman, T. R. 1991, 
     \apj, 371, 684

  \bibitem[1995]{Bl95} Blondin, J. M., \& Woo, J. W. 1995, \apj, 445, 889

  \bibitem[1998]{Bisi98} Bisikalo, D. V., Boyarchuk, A. A., Chechetkin, V. M., 
          Kuznetsov, O. A., \& Molteni, D. 1998, \mnras, 300, 39

  \bibitem[2001]{Boro01} Boroson, B., Kallman, T., Blondin, J. M., \&
      Owen, M. P. 2001, \apj, 550, 919

  \bibitem[1994]{BA94}Boffin, H. M. J. \& Anzer, U. 1994, \aap,  284, 1026  

  \bibitem[1988]{BJ88}Boffin, H. M. J. \& Jorissen, A. 1988, \aap, 205, 155

  \bibitem[1994]{BZ94}Boffin, H. M. J. \& Z\`acs, L. 1994, \aap, 291, 811

  \bibitem[1944]{1944MNRAS.104..273B}Bondi, H., \& Hoyle, F. 1944, 
     \mnras, 114, 195

  \bibitem[1998]{Carq98} Carquillat, J.M., Jorissen, A., Udry, S., \& Ginestet, 
      N. 1998, \aaps, 131, 49

  \bibitem[2000]{Dumm00} Dumm, T., Folini, D., Nussbaumer, H., Schild, H., 
      Schmutz, W., \& Walder, R. 2000, \aap, 354, 1014

  \bibitem[1999]{FoRu2} Foglizzo, T., \& Ruffert, M. 1997, \aap, 320, 342

  \bibitem[1999]{FoRu1} Foglizzo, T., \& Ruffert, M. 1999, \aap, 347, 901

  \bibitem[2001]{2001PThPh.106..729F} Fujiwara, H., Makita, M., Nagae, T., 
     \& Matsuda, T. 2001, Prog. Theor. Phys., 106, 729

  \bibitem[2002]{gaw02} Gawryszczak, A. J., Miko{\l}ajewska, J.,
\& R{\' o}{\. z}yczka, M. 2002, A\&A, 385, 205

  \bibitem[2001]{2001ApJ...558..323H}Hachisu, I., \& Kato, M. 2001, 
     \apj, 558, 323-350

  \bibitem[1939]{HL}Hoyle, F., \& Lyttleton, R. A. 1939, Proc. Camb. Phil. 
     Soc., 35, 405

  \bibitem[1993]{1993ApJ...407L..81K}Kenyon, S J., Livio, M., 
     Miko{\l}ajewska, J., \& Tout, C. A. 1993, \apj, 407, 81

  \bibitem[2000]{2000MNRAS.316..906M} Makita, M., Miyawaki, K., 
     \& Matsuda, T. 2000, \mnras, 316, 906

  \bibitem[1998]{MM98} Mastrodemos, N., \& Morris, M. 1998, \apj, 497, 303

  \bibitem[1987]{1987MNRAS.226..785M} Matsuda, T., Inoue, M., \& Sawada, K. 
     1987, \mnras, 226, 785

  \bibitem[1992]{1992MNRAS.255..183M} Matsuda, T., Ishii, T., Sekino, N., 
     Sawada, K., Shima, E., Livio, M., \& Anzer, U. 1992, \mnras, 255, 183

  \bibitem[2000]{2000Ap&SS.274..259M} Matsuda, T., Makita, M., Fujiwara, H., 
     Nagae, T., Haraguchi, K., Hayashi, E., \& Boffin, H. M. J. 2000, 
     \apss, 274, 259

  \bibitem[1997]{Miko}Miko{\l}ajewska, J. 1997, in 
     ``Physical Processes in Symbiotic Binaries and Related Systems,'' 
     ed. J. Miko{\l}ajewska, (Copernicus Foundation for Polish Astronomy: 
     Warsaw), p. 3

  \bibitem[1999]{1999A&AS..137..473M} M\"urset U., \& Schmid H. M. 1999,
     \aaps, 137, 473

  \bibitem[1991]{Nuss}Nussbaumer, H. 1991, Evolution in Astrophysics, 
     Proc. IUE Symposium, Toulouse, FR, (ESTEC: Noordwijk)

  \bibitem[2000]{Pochan00} Pogorelov, N.V., Ohsugi, Y., \& Matsuda, T. 2000, 
    \mnras, 313, 198

  \bibitem[1996]{Ruffert96} Ruffert, M. 1996, \aap, 311, 817

  \bibitem[1984]{1984MNRAS.206..673S} Sawada, K., Hachisu, I. \& Matsuda, T. 
     1984, \mnras, 206, 673

  \bibitem[1986]{1986MNRAS.221..679S} Sawada, K., Matsuda, T., \& Hachisu, I. 
     1986, \mnras, 221, 679

  \bibitem[1992]{1992PASP..104...87S} Sion, E. M., \& Ready, C. 1992, 
     \pasp, 104, 87

  \bibitem[1994]{1994ApJ...421..261S}Sion, E. M., \& Starrfield, S. G. 1994, 
     \apj, 421, 261

  \bibitem[1975]{1975Ap&SS..33..465S} Sorensen, S.-A., Matsuda, T., 
     \& Sakurai, T. 1975, \apss, 33, 465

  \bibitem[1993]{TJ93} Theuns, T. \& Jorissen, A. 1993, \mnras, 265, 946

  \bibitem[1996]{TBJ96} Theuns, T., Boffin, H. M. J., \& Jorissen, A. 1996, 
     \mnras, 280, 1264

  \bibitem[1997]{Walder97} Walder, R. 1997, 
     in Accretion phenomena and related outflows, 
          Wichramasinghe D.T., Ferrario L., Bicknell G.V. (eds.), IAU
          Colloquium 163, ASP Conf. Ser. No. 121, p. 822

 \bibitem[2000]{Walder00} Walder, R., \& Folini, D. 2000, 
     in Thermal and ionization Aspects of Flows from Hot Stars, 
         Lamers, H., \& Sapar, A. (eds.), ASP Conf. Ser. No. 204, p. 331
\end{thebibliography}
\end{document}